\documentclass[conference]{IEEEtran}
\IEEEoverridecommandlockouts
\usepackage{cite}
\usepackage{amsmath,amssymb,amsfonts}
\usepackage{algorithmic}
\usepackage{graphicx}
\usepackage{textcomp}
\usepackage{xcolor}
\usepackage{dsrm}
\usepackage{hyperref}
\usepackage{svg}
\def\BibTeX{{\rm B\kern-.05em{\sc i\kern-.025em b}\kern-.08em
    T\kern-.1667em\lower.7ex\hbox{E}\kern-.125emX}}
\usepackage{tikz}
\usetikzlibrary{fit, backgrounds, decorations.markings}
\usepackage[caption=false,font=footnotesize]{subfig}
\begin{document}

\title{Navigating the Socio-Technical Complexity Challenge in Quantum Software Ecosystems\\

}

\newcommand{\snc}[1]{\textcolor{blue}{#1}}
\newcommand{\vlad}[1]{\textcolor{olive}{#1}}

\author{\IEEEauthorblockN{Ronja Heikkinen}
\IEEEauthorblockA{\textit{University of Jyväskylä}\\
Jyväskylä, Finland \\
ronja.k.heikkinen@jyu.fi}
\and
\IEEEauthorblockN{Santiago Núñez-Corrales}
\IEEEauthorblockA{\textit{University of Illinois Urbana-Champaign}\\
Urbana, IL, US\\
nunezco2@illinois.edu
}
\and
\IEEEauthorblockN{Vlad Stirbu}
\IEEEauthorblockA{\textit{University of Jyväskylä}\\
Jyväskylä, Finland \\
vlad.a.stirbu@jyu.fi}
}

\maketitle

\begin{abstract}
Quantum computing environments are composed of heterogeneous layers spanning hardware, software development kits, and applications. Practitioners curating these environments face a fragmented and rapidly evolving landscape with few principled guides for navigation. This paper presents a framework for evaluating quantum computing environment choices through a socio-technical lens, developed using Design Science Research methodology. Drawing on the quantum software engineering literature as well as organizational  and socio-technical research, the framework introduces three analytical constructs: gravity wells and their properties, which characterize how certain technologies and structural conditions exert increasing pull on surrounding environment choices, and socio-technical desiderata, which articulate the normative goals against which those pulls can be evaluated. The framework supports practitioners in making deliberate, context-aware environment choices that preserve architectural flexibility and support the evolutionary development of the field. Demonstration and evaluation of the framework is conducted through exemplary cases. The contribution advances both the theory of quantum ecosystems and the practical guidance available to organizations and practitioners navigating the current, evolving field of quantum computing.
\end{abstract}


\section{Introduction}

Computing infrastructures are deeply intertwined with organizational practices, shaping how work is performed and how flexibility emerges in research and industrial settings. In advanced computing domains, such as quantum computing (QC) and high-performance computing (HPC), this relationship becomes more pronounced due to the complexity, scarcity, and heterogeneity of resources. QC, in particular, represents a disruptive shift, not only in computational capabilities but also in how users interact with infrastructure, as access is typically remote, mediated, and evolving alongside the technology itself. As a result, integrating quantum and classical resources introduces new challenges in maintaining usable and adaptable computing environments.

In practice, computing environments are not seamless systems but require continuous adaptation by users to compensate for misalignments between infrastructure and work requirements. In the context of quantum computing, this challenge is amplified by technological immaturity and fragmentation across the software stack. Users face a growing number of choices in tools, frameworks, and execution environments, which creates combinatorial complexity and prevents the formation of stable development routines. This leads to increased mental and operational burden, where users must manage heterogeneous resources, bridge incompatible systems, and absorb the hidden costs of integration. The result is a tension between the need for experimentation in an emerging field and the constraints imposed by limited resources and evolving infrastructure.

To address this gap, we propose a framework for analyzing quantum software environments through a socio-technical lens. The framework introduces the concept of gravity wells to capture how technologies and practices exert structural pull on surrounding choices, and a set of socio-technical desiderata to evaluate whether this pull enables or constrains experimentation and evolution. Developed using Design Science Research (DSR), the framework is demonstrated through three representative execution scenarios spanning cloud-native and HPC-oriented approaches. Rather than prescribing a single solution, the contribution provides a structured method for identifying sources of friction, reasoning about architectural trade-offs, and guiding the design of more evolvable and practitioner-friendly hybrid quantum–classical environments.

The paper is structured as follows. Section~\ref{sec:background} provides the background. Section~\ref{sec:methodology} describes the methodology and the objectives. Section~\ref{sec:design} describes the design of the framework. Section~\ref{sec:demonstration}  presents the use of the framework in representative scenarios, followed by Section~\ref{sec:evaluation}  in which the framework is evaluated. The paper concludes in Section~\ref{sec:conclusion}.

\section{Background and motivation}
\label{sec:background}

Existing research has demonstrated progress toward integrating heterogeneous computing resources \cite{beck-etal-2024-integrating, schulz-etal-2023-accelerating, chazapis-etal-2023-running, manzi-etal-2025-intertwin}. The proposals are often limited to experimental setups or specific institutions, and their applicability in real-world scientific workflows remains constrained. In hybrid QC-HPC execution developer experience continues to be fragmented and vendor-dependent \cite{destefano-etal-2022-towards, beck-etal-2024-integrating, schulz-etal-2023-accelerating}. We have noted a clear research gap: the need for practical, developer-oriented orchestration solutions that enable interoperability across heterogeneous infrastructures while reducing workflow complexity. In essence, structural tension is evident between QC, HPC, cloud, and scientific demands.

\begin{figure}
    \centering
    \includegraphics[width=0.98\linewidth]{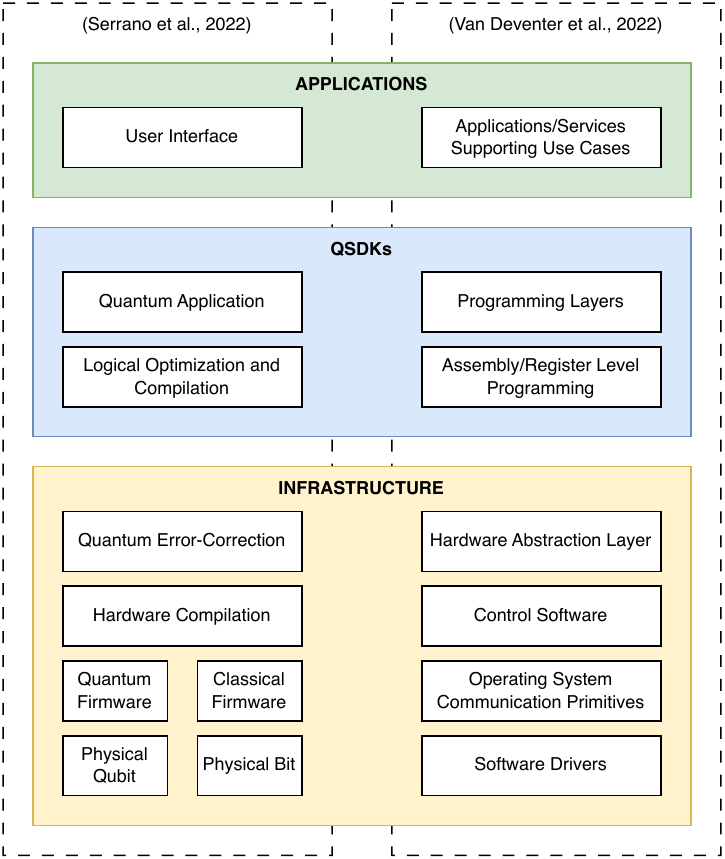}
    \caption{The three layers of the quantum software stack, based on \cite{serrano-etal-2022-quantum,vandeventer2022quantumstandards}}
    \label{fig:stack}
\end{figure}

\subsection{Quantum Software Stack}

The quantum software stack refers to the organized layering of components built on top of hardware (Fig. \ref{fig:stack}). It clarifies that when practitioners speak of fragmentation or interoperability, they are often diagnosing misalignment between specific layers rather than the stack as a whole. \cite{vandeventer2022quantumstandards} and  \cite{serrano-etal-2022-quantum} propose structures for the stack, showing that the most consistent gaps appear precisely at the interfaces between layers.

In the infrastructure level, there exists a growing range of alternatives in terms of proximity of QC to HPC systems. Effectively orchestrating those remote computational resources can significantly enhance quantum software development \cite{kjorveziroski-filiposka-2022-kubernetes, li-zhao-2024-moirai}. Middleware solutions can simplify interactions, decouple quantum logic from underlying systems, and dynamically allocate workloads across QC and HPC resources \cite{serrano-etal-2022-quantum, stirbu-mikkonen-2023-software}.  Efforts such as Quantum Device Interface Layer (QDIL)\footnote{\href{https://www.newquantumera.com/blog/why-i-joined-the-unitary-foundation-and-what-im-planning-to-do/}{https://www.newquantumera.com/blog/why-i-joined-the-unitary-foundation-and-what-im-planning-to-do/}} and Quantum Device Management Interface (QDMI) \cite{wille2024qdmi} seek to maintain homogeneous access across devices while recognizing the need to operate within the specifics of each modality and potentially their integration. 

Above the infrastructure layer, quantum software development kits (QSDKs) operationalize access to quantum hardware and hybrid resources \cite{singh-etal-2024-survey,serrano-etal-2022-quantum,hevia-etal-2022-quantumpath,marosi-etal-2023-toward,murillo-etal-2025-quantum}. Despite the plethora of available SDKs, many languages and frameworks remain research-oriented or vendor-specific, full interoperability across platforms being limited \cite{serrano-etal-2022-quantum,haghparast-etal-2023-quantum,murillo-etal-2025-quantum} and noise sensitivity constraining these tools in the NISQ era \cite{sanchez-alonso-2021-definition,chong-etal-2017-programming}. Within the layered perspective adopted in this background section, QSDKs function as the critical mediation layer. Their maturity directly influences portability, interoperability, and developer experience across hybrid QC-HPC environments.

At the top of the stack, applications materialize quantum and hybrid capabilities into domain-specific workflows. Access to quantum applications is predominantly mediated through cloud-based service models that abstract away hardware ownership and management, such as the as-a-service framework \cite{ahmad-etal-2023-engineering,destefano-etal-2022-towards,nguyen-etal-2024-qfaas,fanzago-etal-2025-cloudveneto}. In scientific computing contexts, applications frequently span heterogeneous infrastructures, combining cloud-native orchestration with remote QC resources and  HPC backends  \cite{manzi-etal-2025-intertwin,ciangottini2025unlocking}, with focus on workflow continuity, scheduling efficiency, and adaptive resource allocation \cite{alam-etal-2025-empirical,schulz-etal-2023-accelerating}. From a layered perspective, the application-level limitations often reflect unresolved architectural tensions in the underlying QC-HPC integration stack.

\subsection{What Has Been Attempted}

Existing research demonstrates substantial progress toward integrating heterogeneous computing resources \cite{beck-etal-2024-integrating, schulz-etal-2023-accelerating, chazapis-etal-2023-running, manzi-etal-2025-intertwin}. Hardware-agnostic approaches \cite{beck-etal-2024-integrating} and shared environments with interoperable tooling \cite{schulz-etal-2023-accelerating} have been gaining attention. \cite{beck-etal-2024-integrating}, \cite{tejedor2026kubernetes}, \cite{seelam2026referencearchitecturequantumcentricsupercomputer} and \cite{marosi-etal-2023-toward} introduce comprehensive frameworks for QC-HPC integration. Critically, \cite{tejedor2026kubernetes} positions Kubernetes\footnote{\url{https://kubernetes.io/}} and Slurm\footnote{\href{https://slurm.schedmd.com}{https://slurm.schedmd.com}} as complementary rather than competing, meaning they can coexist in hybrid HPC deployments, whereas much prior discussion treats Kubernetes and Slurm as mutually exclusive.

Effective integration requires management at two levels: workflows and resources \cite{beck-etal-2024-integrating}. Composable tooling and bridging solutions advance interoperability but fall short of delivering end-to-end, vendor-neutral orchestration suitable for sustained scientific use \cite{chazapis-etal-2023-running, ciangottini2025unlocking, manzi-etal-2025-intertwin}, with scheduling remaining a problem \cite{schulz-etal-2023-accelerating,alam-etal-2025-empirical}. Some offered solutions include Quantum Algorithms as a Service (QAaaS) \cite{destefano-etal-2022-towards}, Qonductor \cite{giortamis-etal-2025-qonductor}, The Bridge Operator \cite{lublinsky-etal-2022-kubernetes}, NVQLink \cite{caldwell-etal-2025-platform}, serverless Quantum Function as a Service (QFaaS) \cite{nguyen-etal-2024-qfaas}, High-Performance Kubernetes (HPK) \cite{chazapis-etal-2023-running}, and Dynamic Kubernetes \cite{decker-kunkel-2025-ephemeral}. Several solutions also address GPU sharing within Kubernetes environments, targeting deep learning and multi-tenant HPC workloads \cite{chung-etal-2025-fine,kim-etal-2021-gpu,li-etal-2023-easyscale}. CloudVeneto uses InterLink\footnote{\url{https://interlink-project.dev/}} to simplify deployment, resource management, and workload offloading \cite{fanzago-etal-2025-cloudveneto}. InterLink allows Kubernetes to execute arbitrary heterogeneous remote resources \cite{manzi-etal-2025-intertwin,ciangottini2025unlocking}. 

\subsection{Portability and Orchestration Problems}

One of the emerging trends in HPC is the integration of QC \cite{arianyan-etal-2025-systematic}, which provides broader access to computational resources that were previously highly specialized and restricted \cite{dai-etal-2025-state, ahmad-etal-2023-engineering}. In this landscape, cloud-native and serverless paradigms \cite{dai-etal-2025-state} as well as containerization technologies such as Docker\footnote{\url{https://www.docker.com/}} and Kubernetes \cite{ahmad-etal-2025-containers} are gaining attention. Actors in HPC seem to notice the benefits of containerization: for example, LUMI supercomputer has its own orchestration platform\footnote{\url{https://docs.lumi-supercomputer.eu/runjobs/lumi-k/getting-started/lumi_k_what_is/}}. Despite these advances, orchestration in multi-cloud and hybrid QC-HPC environments remains underexplored \cite{ahmad-etal-2025-containers}, with existing workflows often forcing developers to manually bridge separate infrastructures \cite{serrano-etal-2022-quantum, stirbu-mikkonen-2023-software}.

The orchestration challenge in classical computing has evolved to meet concrete needs in industry and academia across variety of environments, but the emerging trade-offs from specific demands arising from the translation of workflow specifications into actual workloads reflect market realities, often captured by a power-law like distribution in the adoption of specific technologies. Market data exhibits this: Market Growth Reports \cite{marketgrowthreports2025container} describe Kubernetes as the dominant container platform. QC orchestration is in earlier stage. QPU orchestration decisions tend to be tied to specific institutional efforts rather than emerging from broad market selection. This contrast between the two domains is not merely descriptive — it underscores why direct translation of classical orchestration patterns to the quantum domain is insufficient, while also confirming that established classical tools like Kubernetes represent a sensible starting point.

Across the stack, quantum software engineering functions as the connective tissue. The field suffers from a structural complexity problem due to choices in the levels of the stack, creating a complex multi-to-multi mapping problem: no single solution can dominate across all resource pairings and workflow types, nor should it. The more productive framing is not which solution wins, but whether the ecosystem can support the full range of pairings that scientific and engineering practice demands. What is needed are tools and frameworks that embrace this heterogeneity, without forcing developers into narrow, vendor-defined corridors. 

\subsection{Socio-Technical Problems}

Technology does not exist in a vacuum. Every tool, platform, or system we deploy enters an existing social context. This is the foundational insight of the socio-technical tradition: technological artifacts and social practices are constitutively entangled. As QC matures, it confronts the same challenge that has accompanied every major computational paradigm shift: how do we build environments where the right tools fit together into a whole genuinely greater than the sum of its parts? The following insights apply to QC environments, where the stack spans multiple heterogeneous layers and no single offering covers every level adequately. Choosing a QC environment is therefore not an act of selecting the best tool, but an act of alignment: matching available offerings to the given context.

Gasser's \cite{gasser1986integration} study of routine computer use demonstrates that even in highly successful computing environments, the fit between systems and the work they support is perpetually unstable. Star and Ruhleder \cite{star1994steps} deepen this account by showing that such friction arises from incommensurable contextual levels: infrastructure is never a passive substrate but always a relation, becoming visible precisely in where it breaks down. Orlikowski continues this insight further, explaining that there is no social that is not also material, and no material that is not also social \cite{orlikowski2007sociomaterial}. Friction is therefore not a residual problem to be eliminated by better engineering, but a persistent condition arising from the gap between systems as designed and work as performed. This tradition stands in deliberate contrast to adoption models such as TAM \cite{davis1989perceived, davis1989user}, UTAUT \cite{venkatesh2003utaut}, and UTAUT2 \cite{venkatesh2012utaut2}, which treat adoption as predictable from individual-level beliefs.

\section{Motivation, Methodology and Objective}
\label{sec:methodology}

This paper follows the DSR methodology \cite{tuunanen2024dealing,peffers2007design}. The problem-centered approach allows us to notice that the quantum software environment is composed of interchangeable but poorly interoperable elements, creating combinatorial configuration complexity. Assembling the right elements and configuring the stack becomes a complex problem with many choices that require skills and expertise. As a result, users cannot form stable routines, while choices carry hidden irreversible costs. Additionally, we identify that the problems being solved with QC must be positioned in a socio-technical context, with a short-term objective to lower the configuration combinatorial complexity and a long-term goal to enable resilience and evolvability.

Therefore, during the \textit{Problem identification and motivation} phase, we were able to distill the following research questions:

\begin{itemize}
    \item RQ1: How can orchestration infrastructure be designed to reduce socio-technical friction and combinatorial configuration complexity in hybrid QC–HPC software environments?
    \item RQ2: What design criteria and quality attributes enable such orchestration frameworks to support flexibility, interoperability, and evolvability?
    \item RQ3: How to identify the dominant sources of friction in quantum software ecosystems and how do they impact developer workflows?
\end{itemize}

In the \textit{Objectives of the solution} phase, the goal was to define a framework that supports practitioners in navigating the socio-technical challenges of quantum software environments, with particular focus on reducing combinatorial configuration complexity and associated frictions. The framework was developed during the \textit{Design and development} phase, grounded in socio-technical theory and operationalized through the concepts of technological evaluation criteria for evaluating individual nodes, gravity wells and their properties for finding the problematic nodes, and socio-technical desiderata for evaluating the impact of the gravity wells. In the \textit{Demonstration} phase, the framework was applied to three representative case scenarios to identify gravity wells and analyze their effects on workflow and environment design. The \textit{Evaluation} phase consisted of a critical reflection on the framework’s explanatory and analytical capabilities in these scenarios. Finally, the \textit{Communication} phase reports the results in this paper.

\section{Design and Development}
\label{sec:design}

\subsection{Criteria for technology evaluation}

                \begin{table}[]
                \footnotesize
                \centering
                \caption{Mapping the evaluation criteria to the software stack levels}
                \label{tab:tech-criteria}
                \begin{tabular}{|p{2.4cm}|p{1.4cm}|p{1.4cm}|p{1.4cm}|}\hline
         & \multicolumn{3}{|c|}{\textbf{Software Stack Levels}}\\\hline
                \hline
                      \textbf{Evaluation Criteria} & App& SDKs&Infrastructure\\ \hline
                      Reproducibility &  \cite{ferreira-campos-2025-exploratory,serrano-etal-2022-quantum,murillo-etal-2025-quantum}& \cite{serrano-etal-2022-quantum,murillo-etal-2025-quantum} &\\ \hline
                      Fault-tolerance &   \cite{murillo-etal-2025-quantum} &  \cite{murillo-etal-2025-quantum} &\\ \hline 
                      Interoperability &  \cite{serrano-etal-2022-quantum,ferreira-campos-2025-exploratory}&  \cite{serrano-etal-2022-quantum,ferreira-campos-2025-exploratory,murillo-etal-2025-quantum}&  \cite{serrano-etal-2022-quantum,ferreira-campos-2025-exploratory}\\ \hline
                      Tooling &  \cite{ferreira-campos-2025-exploratory}& &\\ \hline 
                      Transparency &  \cite{serrano-etal-2022-quantum}&  \cite{murillo-etal-2025-quantum,ferreira-campos-2025-exploratory}& \cite{ferreira-campos-2025-exploratory,serrano-etal-2022-quantum}\\ \hline 
                      Usability &  \cite{ferreira-campos-2025-exploratory,murillo-etal-2025-quantum}& &\\ \hline
                       Programmability &  &  \cite{murillo-etal-2025-quantum,serrano-etal-2022-quantum} &  \cite{murillo-etal-2025-quantum,serrano-etal-2022-quantum} \\ \hline
                        Development processes &  &  \cite{serrano-etal-2022-quantum,murillo-etal-2025-quantum}&\\ \hline 
                         Deployment model &  & &  \cite{murillo-etal-2025-quantum}  \\ \hline 
                         Implementation technology & & &  \cite{serrano-etal-2022-quantum} \\ \hline 
                         Use model &  & &  \cite{serrano-etal-2022-quantum,murillo-etal-2025-quantum} \\ \hline 
                         Realiability &  & &  \cite{serrano-etal-2022-quantum,ferreira-campos-2025-exploratory}\\ \hline 
                         Backend offerings &  & &  \cite{serrano-etal-2022-quantum,murillo-etal-2025-quantum} \\ \hline 
                         Computing models &  & &  \cite{serrano-etal-2022-quantum,murillo-etal-2025-quantum} \\ \hline 
                \end{tabular}
            \end{table}

When solving quantum problems, practitioners must curate their own quantum computing environments.  The result is not a simple selection problem but a deeply constrained compositional challenge: each choice at one layer carries implications for other layers. If the totality of available environments were represented as a graph, with nodes corresponding to tools and offerings, and edges representing the technical and organizational relationships between them, the evaluation criteria developed in this framework would serve as a lens for identifying where that graph breaks down or centralizes. The evaluation criteria were   synthesized from three comprehensive works: \cite{murillo-etal-2025-quantum,serrano-etal-2022-quantum,ferreira-campos-2025-exploratory}.  Previous work in \cite{heikkinen-etal-2025-towards} about quantum software developer experience was used as a starting point. The criteria mapped to the levels is available in Table \ref{tab:tech-criteria}.

Given the amount of tools and solutions in this evolving field, \textbf{interoperability} between the offerings is essential. It is achieved through API portability, hybrid workflow support, workflow exportability, pluggable backends, clear differentiation of orchestration from vendor-specific dependencies, and agnosticism in different components \cite{serrano-etal-2022-quantum, ferreira-campos-2025-exploratory}. Vendor lock-in is repeatedly identified as a systemic problem that restricts portability, constrains interoperability, and limits architectural evolution \cite{matthews-2021-get, moller-vuik-2017-impact, murillo-etal-2025-quantum, stirbu-mikkonen-2023-software}. A compounding challenge arises from the fundamental asymmetry between QC and HPC systems: HPC infrastructure is designed for multi-user, low-latency operation, unlike QC \cite{schulz-etal-2023-accelerating}. Integration efforts have revealed particular difficulties around remote hardware control \cite{beck-etal-2024-integrating}. Many HPC environments impose additional access barriers, including vendor lock-in \cite{petrosyan-astsatryan-2022-serverless}. In cloud-native settings, GPU-sharing remains uncommon \cite{chung-etal-2025-fine, kim-etal-2021-gpu}, while scientific computing demand increasingly exceeds the capacity of current isolated infrastructures \cite{ciangottini2025unlocking}. The development of quantum-aware frameworks for hybrid systems, alongside runtime management is widely regarded as a central objective for the field \cite{khan-etal-2023-software}. Cloud-HPC workflows that do not require users to connect directly to HPC systems remain uncommon in everyday practice \cite{chazapis-etal-2023-running}, and existing HPC schedulers lack the flexibility characteristic of cloud-native tooling \cite{petrosyan-2025-scheduling}.

Vendor lock-in undermines the \textbf{reproducibility} that interoperability is intended to enable. In quantum computing, reproducibility can be supported through version control and open-source code \cite{ferreira-campos-2025-exploratory}, as well as the development of reusable quantum software components \cite{serrano-etal-2022-quantum, murillo-etal-2025-quantum}. Programmability, in turn, offers users meaningful control through integrated environments \cite{murillo-etal-2025-quantum}. 

Alongside reproducibility, \textbf{transparency} constitutes a critical concern for the continued evolution of the field \cite{serrano-etal-2022-quantum}. Transparency can be realized through workflow abstractions and debugging support, cost and resource visibility, and the deliberate exposure of conceptual stack layers to the user. Governance and management processes further enable transparent, auditable workflows \cite{serrano-etal-2022-quantum}. Transparency renders workflows more legible, making it possible to switch backends, inspect the translation layers of the stack, and maintain continuity in the stack levels. Release stability and sustained maintenance activity are further prerequisites \cite{murillo-etal-2025-quantum, ferreira-campos-2025-exploratory}. Transparency encompasses visibility into HPC scheduler configuration, monitoring across heterogeneous resources, unified logging, and QPU backend configurations. It is also worth noting that some available solutions are oriented toward the \textbf{fault-tolerant} era of quantum computing, while others are not \cite{murillo-etal-2025-quantum}. \textbf{Reliability} is also an important consideration when evaluating infrastructure choices: vendors differ  in the error-correction mechanisms and reliability assurances they offer \cite{serrano-etal-2022-quantum}, as well as in their approaches to performance efficiency and resource utilization \cite{serrano-etal-2022-quantum, ferreira-campos-2025-exploratory}.

\textbf{Development processes} assume particular significance. The emergence of new hybrid software life cycles \cite{serrano-etal-2022-quantum, murillo-etal-2025-quantum}, the iterative development of hybrid software \cite{murillo-etal-2025-quantum}, and the management of quantum software development projects \cite{serrano-etal-2022-quantum} are all active areas of concern. When managing multi-stage scientific workflows, users must orchestrate computation and data across diverse and heterogeneous resources \cite{zhou-etal-2023-orchestration}, and employ disparate bridging tools that enable job transfer while keeping underlying resources differentiated \cite{chazapis-etal-2023-running, decker-kunkel-2025-ephemeral}. Difficulties with workflow execution and orchestration are commonly reported challenges for users \cite{alam-etal-2025-empirical}. 

A structural limitation of the current landscape is the absence of sufficiently mature \textbf{tooling}, with many existing tools remaining at the prototype stage \cite{hevia-etal-2022-quantumpath, khan-etal-2025-advancing, moller-vuik-2017-impact}. \textbf{Usability} is closely coupled with tooling quality. Adequate documentation and illustrative examples \cite{ferreira-campos-2025-exploratory}, continuous updates \cite{ferreira-campos-2025-exploratory}, user-friendly interfaces \cite{murillo-etal-2025-quantum} and clarity of error feedback all significantly affect practitioners' capacity to work effectively with their chosen tools. QC demands substantial levels of domain-specific skills and expertise \cite{hevia-etal-2022-quantumpath, awan-etal-2022-quantum, juarez-ramirez-etal-2024-skills, moller-vuik-2017-impact, destefano-etal-2022-towards}, and the shortage of technical expertise remains an obstacle in the field \cite{awan-etal-2022-quantum}. 

The \textbf{deployment model} has significant downstream implications for the rest of the environment. The \textbf{implementation technology} constrains which adjacent technologies can be meaningfully integrated. The \textbf{use model} determines the cost structure under which resources are accessed. \textbf{Backend offerings} define the set of providers a practitioner can realistically work with. Finally, for certain use cases, \textbf{computing models}, including distributed computing approaches or multi-tenant support, introduce further constraints and opportunities that must be considered in environment design.

\subsection{Gravity Wells and Their Properties}

\begin{figure}
            \centering
            \includegraphics[width=\linewidth]{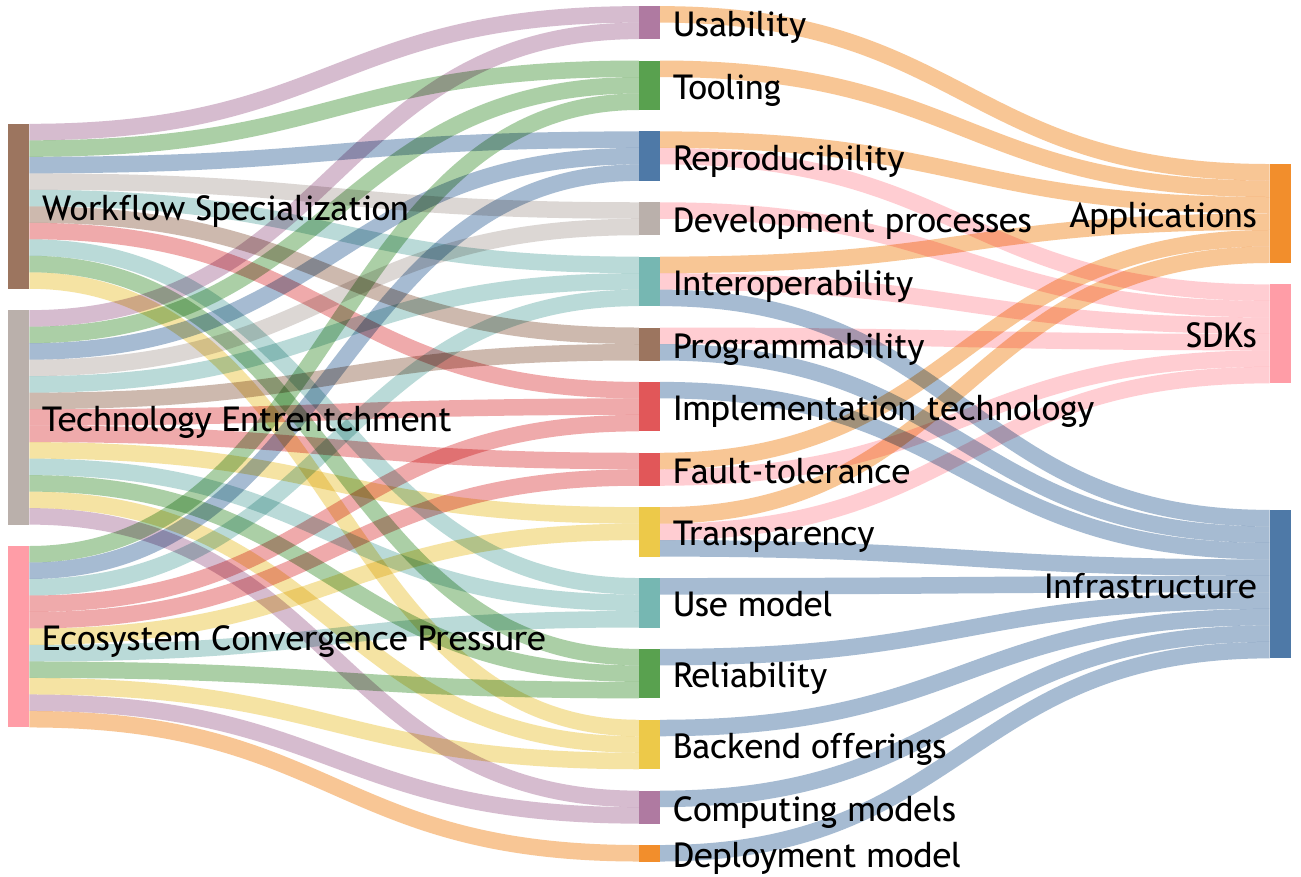}
            \caption{Associations between the gravity well properties, the technology evaluation criteria and the software stack levels (as mapped in Table \ref{tab:tech-criteria})}
            \label{fig:sankey}
        \end{figure}

The concept of a gravity well, as used in this framework, denotes a highly connected partition of the QC environment graph. It captures how certain technologies, practices, or structural conditions, once adopted, exert increasing pull on surrounding choices, shaping which paths through the stack remain practically available. This pull manifests along a spectrum: at one end it enables coherent ecosystem development and productive specialization; at the other it constrains the experimentation space that innovative work requires. This dynamic is reflected in the MAYA principle from design theory, which holds that effective artifacts balance typicality and novelty rather than collapsing into either extreme \cite{hekkert2003most}. Understanding where a given environment sits along this spectrum -- and why -- is the purpose of the gravity well properties analysis that follows. From the technological evaluation criteria, we derive three properties of the gravity wells. These properties help to identify if a partition of a graph is indeed a gravity well. The connections can be seen in Fig.~\ref{fig:sankey}.

\textbf{Workflow Specialization} arises when teams optimize workflows for their immediate operational needs. This process causes knowledge and tooling choices to concentrate around the people who built or routinely use a given system. The spectrum of this  property runs from deep workflow expertise and highly optimized local pipelines at one end, to fragmented development practices, elevated coordination overhead, and reduced transferability at the other. The degree to which this pull opens or closes the surrounding design space depends on the availability of shared orchestration standards and the portability of the chosen bridging solutions.

Infrastructure naturally solidifies as it matures, and tooling attracts users who, through familiarity and accumulated investment, come to depend on it irrespective of its current fit for purpose. The same dynamic applies to processes and methodologies. The spectrum of the \textbf{Technology Entrenchment} runs from stable, high-performance infrastructure at one end, to resistance to architectural evolution and accumulating technical debt at the other. Its effects range from productive stability to structural rigidity depending on the pace of change in the surrounding environment.

At the ecosystem level, certain platforms and toolchains exert gravitational pull that shapes which paths remain available. The spectrum of \textbf{Ecosystem Convergence Pressure} runs from rich, deeply integrated proprietary ecosystems with extensive tooling support at one end, to fragmented, lock-in-prone landscapes where portability and interoperability are structurally constrained at the other. The resulting dependencies range from enabling access to a coherent and well-resourced ecosystem to foreclosing the architectural flexibility that evolving problem contexts may demand.

The three gravity well properties are neither independent nor static. They interact across stack levels, and a practitioner will encounter them simultaneously in different gravity wells. Each property describes a kind of pull, and each pull has a spectrum of possible outcomes depending on the problem context. The purpose of identifying these properties is not to prescribe which ones to avoid, but to make their dynamics legible, so that the paths through the environment graph can be considered with an understanding of what forces are shaping them.

\subsection{Socio-Technical Desiderata}

    \begin{figure*}
        \centering
        \includegraphics[width=\linewidth]{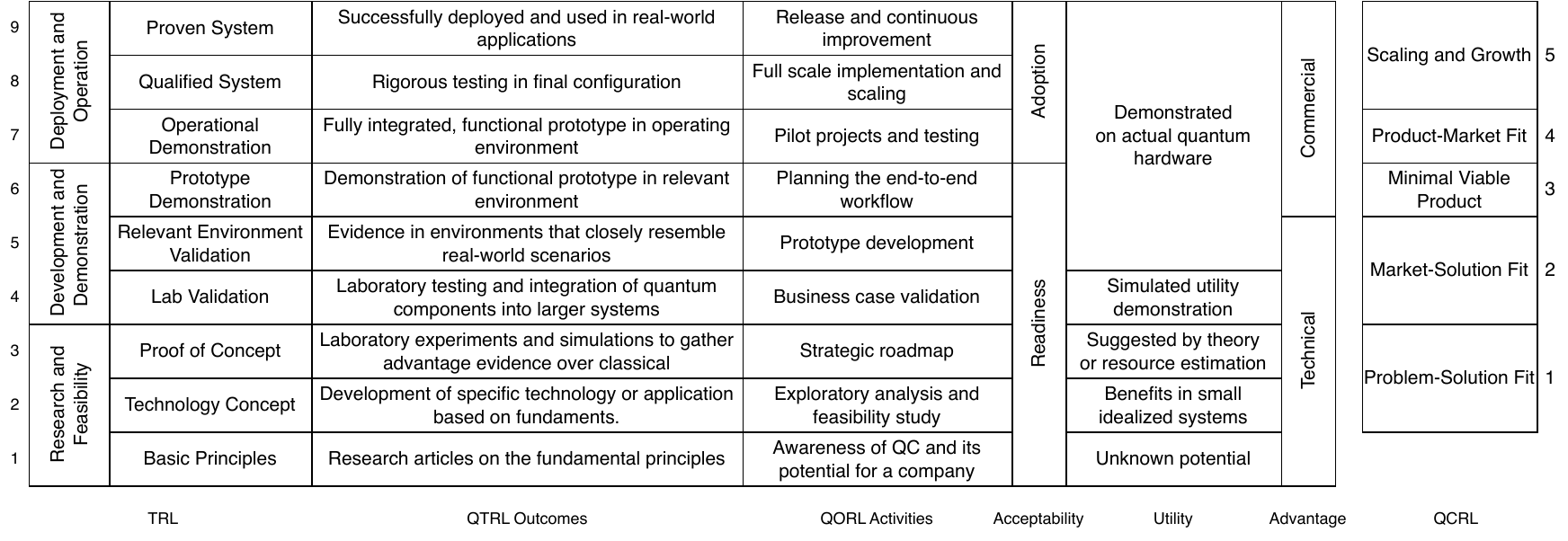}
        \caption{Alignment of readiness levels along different dimensions: general TRL, outcomes corresponding to quantum TRL (QTRL) \cite{purohit2024building}, activities corresponding to quantum organization readiness (QORL) \cite{teitsma2025quantum}, quantum technology acceptability \cite{silberer2023}, quantum application utility \cite{herrmann2023quantumutility}, quantum advantage \cite{bova2023quantumeconomicadvantage}, and quantum commercial readiness (QCRL) \cite{teitsma2025quantum}.}
        \label{fig:trl}
    \end{figure*}

To evaluate whether a given gravity well is productive or obstructive in a particular context, a normative reference point is needed: the socio-technical desiderata developed in this section. They are not prescriptive engineering requirements, but goal-oriented criteria derived from the intersection of the socio-technical tradition reviewed earlier and the technology readiness levels (TRLs) (Fig. \ref{fig:trl}) discussed below. Taken together, they articulate what a well-functioning, evolvable, and practitioner-enabling QC environment should strive toward. When a gravity well does not facilitate progress toward these desiderata, it can be characterized as non-beneficial in that context. This evaluation is context-dependent: the same well may be productive for a team operating at quantum organizational readiness level 3 and obstructive for one operating at level 7. The framework does not resolve that judgment on behalf of the practitioner; instead, it provides a conceptual vocabulary that enables them to make informed, deliberate decisions with a clear understanding of the trade-offs involved.

Quantum technologies qualify as emerging technologies due to their low technological maturity and nascent practical applications \cite{purohit2024building}, with the stack remaining open to fundamental changes at all levels (Fig. \ref{fig:trl}). Mapping the quantum stack onto the \cite{townes2026gtrl} framework, most hardware and software tooling sits in the technical feasibility range, with maturity concentrated only around simulators and classical control infrastructure. This asymmetry means that orchestration solutions must accommodate fundamentally different readiness levels on each side of the QC-HPC integration. \cite{silberer2023} implies that it is a structural constraint on adoption rather than a transitional inconvenience, making developer-oriented orchestration a prerequisite for commercial uptake. Most quantum algorithms of practical interest remain at the theorized-but-invalidated stage of \cite{herrmann2023quantumutility}  framework, creating a reinforcing cycle where gaps in application readiness constrain infrastructure investment. \cite{bova2023quantumeconomicadvantage} sharpen this picture by separating quantum advantage -- outperforming a classical counterpart in accuracy, power, or time \cite{herrmann2023quantumutility} -- from quantum economic advantage, arguing that commercially relevant speedups may arise before classical impossibility is demonstrated, implying that hybrid QC-HPC workflows can be economically justifiable well before fault-tolerant hardware arrives.

As quantum software stacks grow in complexity, monitoring and logging capabilities tend to be added reactively and inconsistently, accumulating blind spots that complicate debugging, reproducibility, and accountability. \textbf{Operational and governance transparency} addresses this tendency by articulating what practitioners must be able to observe and control within their environments. At the operational level, this includes visibility into scheduler configuration, resource allocation decisions, QPU backend parameters, and the translation layers. Unified logging across heterogeneous classical and quantum resources, and monitoring dashboards that expose the state of multi-stage workflows, are practical expressions of this desideratum. At the governance level, transparency requires clear deployment guarantees, reproducible environment configurations, and the ability for practitioners to meaningfully influence the tools and standards. 

A major source of irreversibility in QC environment choices is the accumulation of human capital. When practitioners invest significant effort in learning something, that investment creates switching costs that go beyond the technical. The expertise itself becomes entangled with the tool, making transitions toward more suitable alternatives disproportionately costly. \textbf{Knowledge and skill portability} as a desideratum therefore encompasses both the transferability of conceptual understanding across platforms and the availability of documentation, shared vocabularies, and educational resources that reduce the degree to which expertise is tied to specific implementations. Open standards, common programming abstractions, and cross-platform frameworks all contribute to portability by ensuring that skills acquired in one context remain applicable in others.

Not all constraints are equal: some narrow the solution space temporarily, while others close off architectural paths in ways that are difficult or prohibitively costly to reverse. \textbf{Avoiding irreversible choices and constraints }concerns the identification and, where possible, the avoidance of the latter category. Irreversibility in QC environments arises from several sources. Compartmentalization is the bundling of functionality within proprietary or tightly coupled systems, and reduces the surface area for substitution. Specification occurs when parts of the system are expressed in vendor-specific terms that cannot be straightforwardly translated to alternative platforms. A further dimension concerns the transition from activity space to information space. In action space artifacts, the irreversibility of the underlying tool choices is effectively amplified because the manual labor of the choices is much larger than in information space.

The evolutionary trajectory of a technological field depends in part on its capacity to generate, evaluate, and selectively retain novel approaches. This requires that the environment \textbf{preserves productive failure and allows experimentation for evolution}. In QC, where the maturity landscape is uneven, the preservation of experimentation space is a central concern. Providing practitioners with interoperable, accessible, and composable solutions allows a form of natural selection to operate across the field: approaches that prove durable under real conditions are retained and developed. This requires that individual tool choices do not carry prohibitive exit costs, that abstraction layers remain penetrable when lower-level control is needed, and that the broader ecosystem does not converge prematurely on solutions whose long-term fitness has not yet been demonstrated. 

\subsection{The composed framework}

\begin{figure*}[h!]
    \centering
    \includegraphics[width=0.8\linewidth]{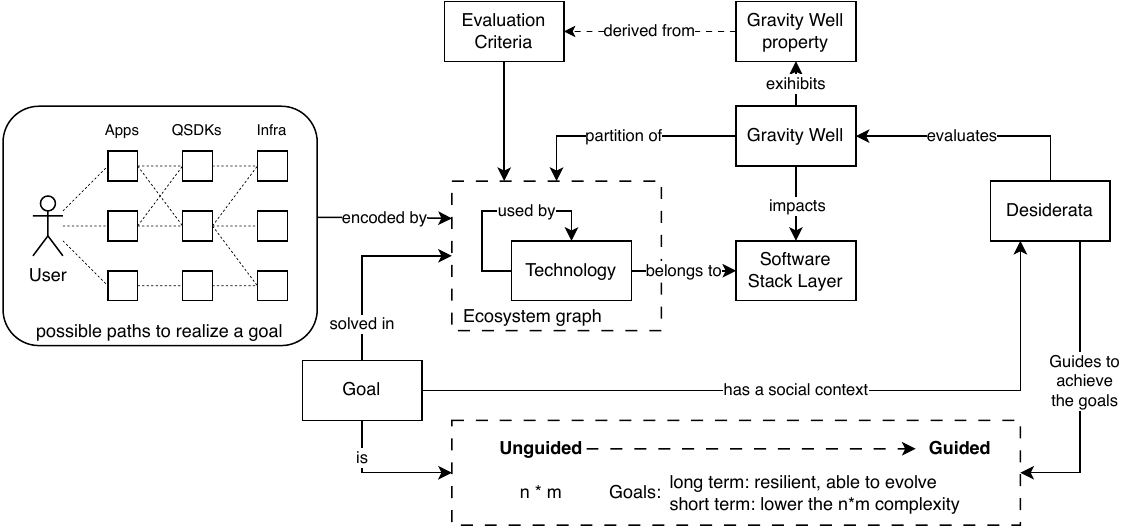}
    \caption{Explained relationships between the framework's core concepts: technological choices, technological ecosystems, evaluation criteria for technology, software stack layers, gravity wells, gravity well properties, socio-technical problems and goals, and the socio-technical desiderata.}
    \label{fig:framework}
\end{figure*}

The framework elements are illustrated in Fig.~\ref{fig:framework}. At the center is the \textit{goal} the user aims to solve using \textit{technologies} drawn from a broader \textit{ecosystem}, not limited to QC. Within this ecosystem, \textit{gravity wells} appear as highly connected regions, that exhibit gravity well \textit{properties}. The technologies within these regions are assessed using the \textit{evaluation criteria}, while their socio-technical impact is interpreted through the \textit{desiderata}. By assigning the technologies to specific \textit{software stack layers}, the impact of each gravity well can be more precisely assessed across the system. 

In the baseline case, progress toward the problem’s goal is \textit{unguided}, exposing the user to maximum complexity. The desiderata provide \textit{guidance}: in the short term, they help reduce this complexity; in the long term, they support resilience and co-evolution with the surrounding technology ecosystem.

\section{Demonstration: Framework Usage in Practice}
\label{sec:demonstration}

In this section, we first apply the framework to three cases representing distinct approaches to QC-HPC workflow execution. Then, we analyze how each supports or constrains experimentation and evolution.

\subsection{Case Studies}

The workload execution scenarios span current QC–HPC integration approaches. Each case is described via its execution model, tracing how a quantum workload moves from the developer environment through scheduling and resource allocation to execution. These cases were chosen in order to evaluate the diversity of execution models for QC-HPC integration and to see the coverage of gravity well properties.

\subsubsection{Qubernetes}

In the Qubernetes execution model \cite{stirbu-etal-2024-qubernetes}, the user initiates execution via the \texttt{q8sctl} toolkit\footnote{\href{https://github.com/qubernetes-dev/q8s-kernel}{https://github.com/qubernetes-dev/q8s-kernel}}, which packages the quantum workload, developed with the user's QSDK of choice, together with its dependencies into a container image. This ensures a reproducible and portable execution environment across targets. The resulting container is submitted to the Kubernetes cluster as a job, where the scheduler assigns it to an appropriate execution target represented as a node. These nodes encapsulate available compute capabilities, including classical resources (CPU, GPU) for simulation and gateways to remote QPU backends for quantum execution. Beyond native cluster resources, Qubernetes integrates HPC environments through mechanisms such as interLink \cite{ciangottini2025unlocking}, exposing them as virtual nodes. This enables HPC resources to participate in the same scheduling and execution model as cloud-native resources, preserving a unified abstraction throughout the execution path.

\subsubsection{Hybrid circuit-cutting workflow}

In this execution model~\cite{tejedor2026kubernetes}, the user initiates the execution path by submitting the workload as an Argo Workflow\footnote{\href{https://argo-cd.readthedocs.io/}{https://argo-cd.readthedocs.io/}}, which represents the multi-stage QC-HPC computation as a declarative pipeline. The workload is containerized, ensuring reproducibility and portability across execution environments. Once deployed to the Kubernetes cluster, the workflow engine orchestrates execution: the Qiskit circuit is partitioned into smaller independent tasks, and Argo manages their dependencies and execution order across the pipeline. The cluster capabilities are extended with Kueue\footnote{\href{https://kueue.sigs.k8s.io}{https://kueue.sigs.k8s.io}}, which provides a unified queuing and scheduling layer for heterogeneous resources. Through this mechanism, nodes exposing CPU, GPU, and QPU capabilities, whether on-premises or accessed remotely, are treated within a single orchestration domain. This allows workloads to be dynamically scheduled to appropriate execution targets based on availability and policy constraints, while maintaining a consistent execution model. In this way, the integration bridges traditional HPC resource management with cloud-native orchestration, completing the execution path through a unified control plane that abstracts underlying infrastructure diversity.

\subsubsection{HPC}

In the HPC execution model, the execution path is initiated by preparing the workload in the developer environment and staging it to the HPC system using standard data transfer mechanisms, such as secure copy or shared file systems. The user then submits the workload through the native HPC scheduler (e.g., Slurm), specifying resource requirements, execution parameters, and the program entry point. The scheduler manages the lifecycle of the job by placing it in a queue and dispatching it to allocated compute nodes once resources become available. These nodes provide classical computation capabilities, including CPUs and GPUs, which are used for simulation, data preparation, and post-processing. Within the execution phase, parallel workloads may leverage MPI to coordinate distributed computation across multiple nodes, enabling scalable simulation and data processing. The interactions with quantum hardware could be handled through the QDMI interface \cite{wille2024qdmi}. Recent work on QC-HPC integration has shown that not only does this pattern translate well when QPUs are present, but that it also enables diversity of execution patterns to serve various needs depending on performance constraints \cite{cipollini2026three,pehlivanoglu2026qurator}.

\subsection{Identifying the gravity wells}

\begin{figure*}
    \centering
    \includegraphics[width=.8\linewidth]{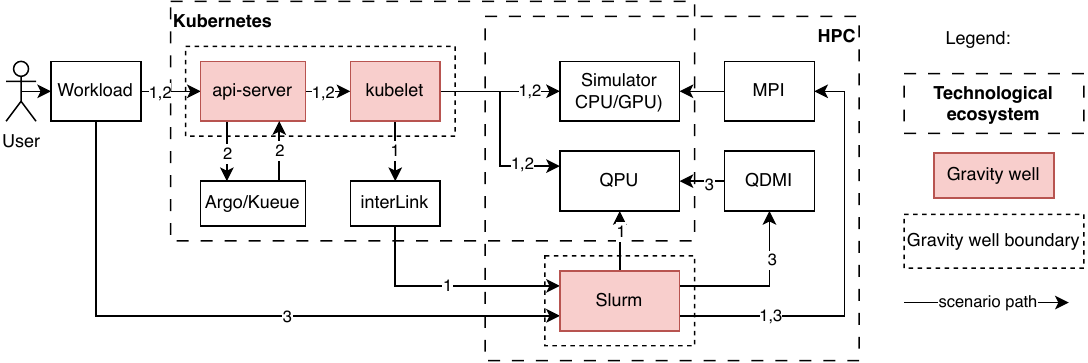}
    \caption{Applying the framework in the selected scenarios. Workloads remain the centerpiece to be realized by hybrid infrastructure. Identified gravity wells determine the choice complexity cost required to marshall infrastructure elements into workflow realizations. HPC-related elements introduce new tensions in the presence of QPUs during technology selection as a function of their TRL.}
    \label{fig:demo-cases-paths}
\end{figure*}

The execution path graphs from the above case studies are extracted and merged into a single comparative view, depicted in Fig.~\ref{fig:demo-cases-paths}. Two components emerge as structural gravity wells, by virtue of the density of connections passing through them: the \texttt{api-server} and \texttt{kubelet} pairing in the Kubernetes-based cases, and \texttt{Slurm} across all three. Both components sit at the convergence of multiple workflow paths, meaning that architectural decisions made with respect to these components propagate across a wide surface of the environment. Their centrality in the graph is not coincidental: it reflects the current state of QC-HPC integration, in which practitioners must route quantum workloads through infrastructure originally designed for classical computation.

The gravity wells' properties are visible in all cases, although with different degree of prominence. The \textit{workflow specialization} is most clearly visible on Slurm, as the primary interface to utilize HPC resources. Its use across all cases reflects the lack of comparable alternatives. Each case integrates it differently, producing workflows tailored to specific needs rather than converging to a shared abstraction. This specialization is structurally characteristic of an emerging field, leading to coordination overhead and concentrated expertise, as noted in the gravity well properties analysis.

The \textit{technology entrenchment} gravity well property is evident in the prominent role of both Kubernetes and Slurm across the cases. Both technologies are mature, widely adopted, and operationally well-understood within their respective domains. Their selection reflects the practical logic of building on established infrastructure. At the same time, this reliance shapes the integration approaches available to practitioners: the Kubernetes-based cases in particular represent an effort to extend cloud-native orchestration semantics into the HPC domain, effectively using Kubernetes as a stable substrate around which novel QC-HPC integration is constructed. This is a productive form of entrenchment insofar as it leverages existing operational knowledge, but it also means that the architectural assumptions embedded in Kubernetes and Slurm propagate into the quantum integration layer, potentially constraining how that layer can evolve.
    
The \textit{ecosystem convergence pressure} gravity well property is present but operates differently across the cases. Kubernetes functions less as a locking ecosystem actor and more as an intermediate orchestration layer that, by design, abstracts the diversity of underlying resources. Its role in the case studies 1 and 2 is closer to that of a neutral integration substrate than a proprietary ecosystem. Slurm, by contrast, operates as a more direct convergence point: its dominance in HPC scheduling means that integration with HPC resources is effectively synonymous with integration with Slurm in the current landscape, regardless of whether practitioners would prefer an alternative. The broader ecosystem convergence pressure -- toward Qiskit as the dominant QSDK, toward the circuit paradigm as the primary abstraction model, and toward specific vendor backends -- is present in the background of all three cases but is not the primary structural feature that the graph makes visible.

\subsection{The gravity wells from the desiderata perspective}

Each identified gravity well contains aspects that support progress toward at least some of the desiderata, while simultaneously introducing tensions with others (Table \ref{tab:case-specific-wells}). No gravity well is uniformly enabling or uniformly constraining across all four dimensions, a finding that is consistent with the spectrum-based framing established in the gravity well analysis and that underscores the context-dependence of any evaluation.

\begin{table*}[ht]
    \footnotesize
    \centering
    \caption{Analysis of the case studies' identified gravity wells using the socio-technical desiderata.}
    \label{tab:case-specific-wells}
    \begin{tabular}{|p{0.8cm}|p{3.7cm}|p{3.7cm}|p{3.7cm}|p{3.7cm}|} \hline 
         & \textbf{Operational and Governance Transparency} & \textbf{Knowledge and Skill Portability} & \textbf{Avoiding Irreversible Choices and Constraints} & \textbf{Preserving Productive Failure and Allowing Experimentation for Evolution} \\ \hline \hline
         \textbf{API Server} & A unified control plane, providing structured visibility & Wide community and adoption & Declarative and extension-based architecture & Extendability and open governance model \\ \hline
         \textbf{Kubelet} & Nodel-level resource states and status, offering unified observability & Well-documented and widely used & Reduces coupling to backends through boundary between control and execution & Node abstractions allow new execution targets, offers virtual nodes without restructuring the broader workflow \\ \hline \hline
         \textbf{Slurm} & Job-level, established logging & Widely used & Configurations directly for HPC & Stable and reliable executions in its domain \\ \hline
    \end{tabular}
\end{table*}

The api-server and kubelet together form a cluster of open, composable dependencies that perform well against transparency, skill portability, and experimentation, but whose contribution to avoiding irreversible constraints depends on the quality of the abstraction maintained at the node boundary. Slurm, by contrast, offers operational maturity and domain-specific transparency within the HPC context, but performs more weakly against portability and the avoidance of irreversible constraints, particularly as quantum workload requirements begin to diverge from the classical batch processing assumptions that Slurm was designed to serve \cite{cacheiro2025qmiotightlyintegratedhybrid}. Across both gravity wells, the desideratum of preserving productive experimentation emerges as the most nuanced to evaluate: each well simultaneously supports and constrains experimentation, depending on whether one is experimenting within the well's operational domain or attempting to integrate across domain boundaries.

The two gravity wells are structurally distinct in character. The Kubernetes api-server is a deliberately extensible and vendor-neutral component as an open-source project with broad community participation. Its design as a general-purpose declarative control plane means that it can accommodate heterogeneous resource types and integration patterns without requiring practitioners to commit to a specific vendor ecosystem. Critically, the tools that extend it in case study~2 operate as additive layers that augment the API server's capabilities without introducing new proprietary constraints. They conform to the Kubernetes extension model, meaning that their adoption does not narrow the architectural choices available downstream. The api-server therefore functions as a high-versatility gravity well: its pull is real and its centrality is significant, but the dependencies it introduces are relatively open and composable.

Slurm, by contrast, is a domain-specific scheduler optimized for the static, tightly managed resource environments of traditional HPC infrastructure. Its dominance in HPC job scheduling means that, in practice, HPC integration is largely synonymous with Slurm integration — regardless the practitioner's preferences.  Unlike the Kubernetes api-server, Slurm does not offer the same degree of architectural openness or community-driven extensibility, and the workflow configurations built around it tend to be more tightly coupled to its scheduling semantics. The gravity well it constitutes is therefore more constraining in character: the pull it exerts tends to concentrate workflow specialization around its specific abstractions and operational assumptions, with fewer composable escape routes available to practitioners who wish to evolve their environments.

\section{Evaluation and Discussion}
\label{sec:evaluation}

In this section, we revisit the research questions, reflect on the framework’s performance in the case studies, and outline the study’s threats to validity.

\subsection{Research questions revisited}

To conclude the contributions of this article, we will provide explicit answers to our research questions:

\begin{itemize}
    \item RQ1: The orchestration infrastructure can be designed by using the proposed framework: identifying gravity wells and evaluating them through the socio-technical desiderata.
    \item RQ2: We identified four attributes: (I) operational and governance transparency, (II) knowledge and skill portability, (III) avoiding irreversible choices and constraints, and (IV) preserving productive failure and allowing experimentation for evolution.
    \item RQ3: Through this study, we concluded that the highly connected points in the technological ecosystem graphs can be gravity wells if they affect workflow specialization, technology entrenchment, or ecosystem convergence.
\end{itemize}

\subsection{Framework Reflection}

The demonstration reveals that the application of the framework can be organized into three sequential phases, corresponding to its three core concepts. In the first phase, the target environment is examined and represented as a technological and/or architectural graph. This graph is then evaluated against the technological criteria, with particular attention paid to heavily connected nodes. In the second phase, potential problem points identified within the graph serve as the basis for recognizing gravity wells. Once the gravity wells have been identified -- and their existence and relative prominence assessed by the proposed gravity well properties -- the third phase begins: evaluating their impact using the desiderata.

This evaluation determines whether the heavily weighted nodes that give rise to the gravity wells genuinely advance the field by creating conditions that enable the desiderata to be satisfied. Taken together, the three phases allow the framework to support more informed and rationally grounded decision-making in the design of technological environments within a given context. In a broad sense, the contribution of this paper, the framework, is not an evaluation of QC-HPC ecosystems but rather a guide on how to evaluate these ecosystems.

The application of the framework to the three cases demonstrates both its analytical reach and its current limitations. On the productive side, the gravity well construct with the three properties proved effective at surfacing structural features of the environment that would not be visible through a purely feature-based evaluation of individual tools. The identification of the API server and Slurm as the two densest nodes in the unified technological graph, and the subsequent analysis of how their structural characters differ, generated insights about architectural risk and evolutionary potential that are not captured by asking simply whether a tool is mature or interoperable. The desiderata similarly proved useful as a multi-dimensional evaluative lens: by requiring the analyst to assess each gravity well against four distinct normative goals rather than a single axis, the framework resists the tendency toward reductive good or bad judgments that would obscure the context-dependence of environment choices.
        
At the same time, the demonstration reveals areas where the framework would benefit from further development. The construction of the technological graph currently relies on the analyst's familiarity with the cases, and a more systematic method for identifying nodes, edges, and gravity wells from documented architectures would increase the framework's reproducibility and reduce its dependence on tacit knowledge. The desiderata evaluations in the table are necessarily qualitative, and the framework does not yet provide guidance on how to weight competing desiderata when they point in different directions. Finally, the three cases examined here represent a specific slice of the QC-HPC integration landscape. Further validation across a broader range of environment configurations would strengthen the generalizability of the framework's findings.

\subsection{Threats to validity}

Several threats to validity, according to \cite{wohlin2012experimentation}, are identified in the context of this work. Regarding conclusion validity, the framework was not subjected to systematic empirical evaluation. Moreover, the evaluation procedure depends on practitioner expertise regarding the system under consideration, rendering the assessment inherently subjective and potentially difficult to reproduce. Internal validity is threatened for analogous reasons. With respect to construct validity, the concepts and criteria comprising the framework represent a first iteration, remain incomplete, and continue to be under development, and thus may not yet fully or accurately capture the underlying theoretical constructs they are intended to operationalize. Finally, external validity is threatened by the fact that the framework was designed with the QC-HPC context specifically in mind.

\section{Conclusions and Future Work}
\label{sec:conclusion}

QC-HPC integration is a socio-technical problem in the broad sense. By characterizing it in this manner and abstracting aspects of the stack to those relevant to individual component choices, we obtain a picture of the complexity and possible reasons for the emergence of frictions across the stack. These partition the space of choices into those creating advantageous and disadvantageous outcomes as we approach gravity wells. The socio-technical desiderata helps us understand the impact of these gravity wells. The framework as a whole guides the evaluation of technological environments. After developing the framework, we demonstrated it in three different use cases: Qubernetes, CERN, and HPC. We applied the framework to these cases and saw the benefits of using it. None of the gravity wells seem to be uniformly enabling or constraining, and the desiderata appear to guide the technological decisions towards enabling a future of evolving through experimentation.

Future work will proceed along four principal directions. First, a systematic survey of existing software solutions will be conducted to characterize the stack. Second, a broader understanding of the scope and interdependencies among gravity wells will be developed. Third, the framework itself will be refined and extended, for instance, by applying graph theory and developing scoring methods for nodes and paths, incorporating TRLs in those methods. Fourth, the framework will be applied to real-world case studies.

\newpage

\section*{Acknowledgments}

This work has been supported by Business Finland projects EM4QS (155/31/2024) and the Research Council of Finland project DEQSE (349945). S. Núñez-Corrales thanks the National Center for Supercomputing Applications for partial funding support via by the Leadership-Class Compute Facility project (NSF \#2323116).

\bibliographystyle{IEEEtran}
\bibliography{bibliography}

\begin{thebibliography}{10}
\providecommand{\url}[1]{#1}
\csname url@samestyle\endcsname
\providecommand{\newblock}{\relax}
\providecommand{\bibinfo}[2]{#2}
\providecommand{\BIBentrySTDinterwordspacing}{\spaceskip=0pt\relax}
\providecommand{\BIBentryALTinterwordstretchfactor}{4}
\providecommand{\BIBentryALTinterwordspacing}{\spaceskip=\fontdimen2\font plus
\BIBentryALTinterwordstretchfactor\fontdimen3\font minus
  \fontdimen4\font\relax}
\providecommand{\BIBforeignlanguage}[2]{{%
\expandafter\ifx\csname l@#1\endcsname\relax
\typeout{** WARNING: IEEEtran.bst: No hyphenation pattern has been}%
\typeout{** loaded for the language `#1'. Using the pattern for}%
\typeout{** the default language instead.}%
\else
\language=\csname l@#1\endcsname
\fi
#2}}
\providecommand{\BIBdecl}{\relax}
\BIBdecl

\bibitem{beck-etal-2024-integrating}
T.~Beck, A.~Baroni, R.~Bennink, G.~Buchs, E.~A.~C. P{\'e}rez, M.~Eisenbach,
  R.~F. da~Silva, M.~G. Meena, K.~Gottiparthi, P.~Groszkowski \emph{et~al.},
  ``Integrating quantum computing resources into scientific hpc ecosystems,''
  \emph{Future Generation Computer Systems}, vol. 161, pp. 11--25, 2024.

\bibitem{schulz-etal-2023-accelerating}
M.~Schulz, M.~Ruefenacht, D.~Kranzlm{\"u}ller, and L.~B. Schulz, ``Accelerating
  hpc with quantum computing: It is a software challenge too,'' \emph{Computing
  in Science \& Engineering}, vol.~24, no.~4, pp. 60--64, 2023.

\bibitem{chazapis-etal-2023-running}
A.~Chazapis, F.~Nikolaidis, M.~Marazakis, and A.~Bilas, ``Running kubernetes
  workloads on hpc,'' in \emph{International Conference on High Performance
  Computing}.\hskip 1em plus 0.5em minus 0.4em\relax Springer, 2023, pp.
  181--192.

\bibitem{manzi-etal-2025-intertwin}
A.~Manzi, R.~Bardaji, I.~Rodero, G.~Molt{\'o}, S.~Fiore, I.~Campos, D.~Elia,
  F.~Sarandrea, A.~P. Millar, D.~Spiga \emph{et~al.}, ``intertwin: Advancing
  scientific digital twins through ai, federated computing and data,''
  \emph{Future Generation Computer Systems}, p. 108312, 2025.

\bibitem{destefano-etal-2022-towards}
M.~De~Stefano, D.~Di~Nucci, F.~Palomba, D.~Taibi, and A.~De~Lucia, ``Towards
  quantum-algorithms-as-a-service,'' in \emph{Proceedings of the 1st
  International Workshop on Quantum Programming for Software Engineering},
  2022, pp. 7--10.

\bibitem{serrano-etal-2022-quantum}
M.~A. Serrano, J.~A. Cruz-Lemus, R.~Perez-Castillo, and M.~Piattini, ``Quantum
  software components and platforms: Overview and quality assessment,''
  \emph{ACM Computing Surveys}, vol.~55, no.~8, pp. 1--31, 2022.

\bibitem{vandeventer2022quantumstandards}
O.~van Deventer, N.~Spethmann, M.~Loeffler, M.~Amoretti, R.~van~den Brink,
  N.~Bruno, P.~Comi, N.~Farrugia, M.~Gramegna, A.~Jenet, B.~Kassenberg,
  W.~Kozlowski, T.~L\"{a}nger, T.~Lindstrom, V.~Martin, N.~Neumann,
  H.~Papadopoulos, S.~Pascazio, M.~Peev, R.~Pitwon, M.~A. Rol, P.~Traina,
  P.~Venderbosch, and F.~K. Wilhelm-Mauch, ``Towards european standards for
  quantum technologies,'' \emph{EPJ Quantum Technology}, vol.~9, p.~33, 2022.

\bibitem{kjorveziroski-filiposka-2022-kubernetes}
V.~Kjorveziroski and S.~Filiposka, ``Kubernetes distributions for the edge:
  serverless performance evaluation,'' \emph{The Journal of Supercomputing},
  vol.~78, no.~11, pp. 13\,728--13\,755, 2022.

\bibitem{li-zhao-2024-moirai}
T.~Li and Z.~Zhao, ``Moirai: Optimizing quantum serverless function
  orchestration via device allocation and circuit deployment,'' in \emph{2024
  IEEE International Conference on Web Services (ICWS)}.\hskip 1em plus 0.5em
  minus 0.4em\relax IEEE, 2024, pp. 707--717.

\bibitem{stirbu-mikkonen-2023-software}
V.~Stirbu and T.~Mikkonen, ``Software architecture challenges in integrating
  hybrid classical-quantum systems,'' in \emph{2023 IEEE International
  Conference on Quantum Computing and Engineering (QCE)}, vol.~2.\hskip 1em
  plus 0.5em minus 0.4em\relax IEEE, 2023, pp. 203--204.

\bibitem{wille2024qdmi}
R.~Wille, L.~Schmid, Y.~Stade, J.~Echavarria, M.~Schulz, L.~Schulz, and
  L.~Burgholzer, ``Qdmi-quantum device management interface: Hardware-software
  interface for the munich quantum software stack,'' in \emph{2024 IEEE
  International Conference on Quantum Computing and Engineering (QCE)},
  vol.~2.\hskip 1em plus 0.5em minus 0.4em\relax IEEE, 2024, pp. 573--574.

\bibitem{singh-etal-2024-survey}
P.~Singh, R.~Dasgupta, A.~Singh, H.~Pandey, V.~Hassija, V.~Chamola, and
  B.~Sikdar, ``A survey on available tools and technologies enabling quantum
  computing,'' \emph{IEEE access}, vol.~12, pp. 57\,974--57\,991, 2024.

\bibitem{hevia-etal-2022-quantumpath}
J.~L. Hevia, G.~Peterssen, and M.~Piattini, ``Quantumpath: A quantum software
  development platform,'' \emph{Software: Practice and Experience}, vol.~52,
  no.~6, pp. 1517--1530, 2022.

\bibitem{marosi-etal-2023-toward}
A.~C. Marosi, A.~Farkas, T.~M{\'a}ray, and R.~Lovas, ``Toward a quantum-science
  gateway: A hybrid reference architecture facilitating quantum computing
  capabilities for cloud utilization,'' \emph{IEEE access}, vol.~11, pp.
  143\,913--143\,924, 2023.

\bibitem{murillo-etal-2025-quantum}
J.~M. Murillo, J.~Garcia-Alonso, E.~Moguel, J.~Barzen, F.~Leymann, S.~Ali,
  T.~Yue, P.~Arcaini, R.~P{\'e}rez-Castillo, I.~Garc{\'\i}a-Rodr{\'\i}guez~de
  Guzm{\'a}n \emph{et~al.}, ``Quantum software engineering: Roadmap and
  challenges ahead,'' \emph{ACM Transactions on Software Engineering and
  Methodology}, vol.~34, no.~5, pp. 1--48, 2025.

\bibitem{haghparast-etal-2023-quantum}
M.~Haghparast, T.~Mikkonen, J.~K. Nurminen, and V.~Stirbu, ``Quantum software
  engineering challenges from developers' perspective: Mapping research
  challenges to the proposed workflow model,'' in \emph{2023 IEEE International
  Conference on Quantum Computing and Engineering (QCE)}, vol.~2.\hskip 1em
  plus 0.5em minus 0.4em\relax IEEE, 2023, pp. 173--176.

\bibitem{sanchez-alonso-2021-definition}
P.~S{\'a}nchez and D.~Alonso, ``On the definition of quantum programming
  modules,'' \emph{Applied Sciences}, vol.~11, no.~13, p. 5843, 2021.

\bibitem{chong-etal-2017-programming}
F.~T. Chong, D.~Franklin, and M.~Martonosi, ``Programming languages and
  compiler design for realistic quantum hardware,'' \emph{Nature}, vol. 549,
  no. 7671, pp. 180--187, 2017.

\bibitem{ahmad-etal-2023-engineering}
A.~Ahmad, M.~Waseem, P.~Liang, M.~Fehmideh, A.~A. Khan, D.~G. Reichelt, and
  T.~Mikkonen, ``Engineering software systems for quantum computing as a
  service: A mapping study,'' \emph{arXiv preprint arXiv:2303.14713}, 2023.

\bibitem{nguyen-etal-2024-qfaas}
H.~T. Nguyen, M.~Usman, and R.~Buyya, ``Qfaas: A serverless
  function-as-a-service framework for quantum computing,'' \emph{Future
  Generation Computer Systems}, vol. 154, pp. 281--300, 2024.

\bibitem{fanzago-etal-2025-cloudveneto}
F.~Fanzago, D.~Lupu, D.~Marcato, M.~Sgaravatto, S.~Traldi, A.~Troja,
  M.~Verlato, and L.~Zangrando, ``The cloudveneto’s container-as-a-service
  ecosystem,'' in \emph{EPJ Web of Conferences}, vol. 337.\hskip 1em plus 0.5em
  minus 0.4em\relax EDP Sciences, 2025, p. 01250.

\bibitem{ciangottini2025unlocking}
D.~Ciangottini, D.~Spiga, A.~S. Memon, A.~Manzi, A.~Filipcic, A.~Troja,
  F.~Fanzago, G.~Bianchini, M.~Sgaravatto, T.~Prica \emph{et~al.}, ``Unlocking
  the compute continuum: scaling out from cloud to hpc and htc resources,'' in
  \emph{EPJ Web of Conferences}, vol. 337.\hskip 1em plus 0.5em minus
  0.4em\relax EDP Sciences, 2025, p. 01296.

\bibitem{alam-etal-2025-empirical}
K.~Alam, B.~Roy, C.~K. Roy, and K.~Mittal, ``An empirical investigation on the
  challenges in scientific workflow systems development,'' \emph{Empirical
  Software Engineering}, vol.~30, no.~5, p. 151, 2025.

\bibitem{tejedor2026kubernetes}
\BIBentryALTinterwordspacing
M.~Tejedor, M.~Grossi, C.~Tüysüz, R.~Rocha, and S.~Vallecorsa,
  ``Kubernetes-orchestrated hybrid quantum--classical workflows,'' 2026.
  [Online]. Available: \url{https://arxiv.org/abs/2603.24206}
\BIBentrySTDinterwordspacing

\bibitem{seelam2026referencearchitecturequantumcentricsupercomputer}
\BIBentryALTinterwordspacing
S.~Seelam, J.~M. Chow, A.~Córcoles, S.~Sheldon, T.~Mittal, A.~Kandala,
  S.~Dague, I.~Hincks, H.~Horii, B.~Johnson, M.~Le, H.~Jamjoom, and J.~M.
  Gambetta, ``Reference architecture of a quantum-centric supercomputer,''
  2026. [Online]. Available: \url{https://arxiv.org/abs/2603.10970}
\BIBentrySTDinterwordspacing

\bibitem{giortamis-etal-2025-qonductor}
\BIBentryALTinterwordspacing
E.~Giortamis, F.~Romao, N.~Tornow, D.~Lugovoy, and P.~Bhatotia, ``Qonductor: A
  cloud orchestrator for quantum computing,'' in \emph{Proceedings of the
  International Conference for High Performance Computing, Networking, Storage
  and Analysis}, ser. SC '25.\hskip 1em plus 0.5em minus 0.4em\relax New York,
  NY, USA: Association for Computing Machinery, 2025, p. 728–745. [Online].
  Available: \url{https://doi.org/10.1145/3712285.3759785}
\BIBentrySTDinterwordspacing

\bibitem{lublinsky-etal-2022-kubernetes}
B.~Lublinsky, E.~Jennings, and V.~Spi{\v{s}}akov{\'a}, ``A kubernetes'
  bridge'operator between cloud and external resources,'' \emph{arXiv preprint
  arXiv:2207.02531}, 2022.

\bibitem{caldwell-etal-2025-platform}
\BIBentryALTinterwordspacing
S.~A. Caldwell, M.~Khazraee, E.~Agostini, T.~Lassiter, C.~Simpson, O.~Kahalon,
  M.~Kanuri, J.-S. Kim, S.~Stanwyck, M.~Li, J.~Olle, C.~Chamberland, B.~Howe,
  B.~Schmitt, J.~G. Lietz, A.~McCaskey, J.~Ye, A.~Li, A.~B. Magann, C.~I.
  Ostrove, K.~Rudinger, R.~Blume-Kohout, K.~Young, N.~E. Miller, Y.~Xu,
  G.~Huang, I.~Siddiqi, J.~Lange, C.~Zimmer, and T.~Humble, ``Platform
  architecture for tight coupling of high-performance computing with quantum
  processors,'' 2025. [Online]. Available:
  \url{https://arxiv.org/abs/2510.25213}
\BIBentrySTDinterwordspacing

\bibitem{decker-kunkel-2025-ephemeral}
J.~Decker and J.~Kunkel, ``Ephemeral kubernetes: dynamically deleting and
  recreating clusters using warewulf,'' \emph{The Journal of Supercomputing},
  vol.~81, no.~16, pp. 1--28, 2025.

\bibitem{chung-etal-2025-fine}
W.-C. Chung, J.-S. Tong, and Z.-H. Chen, ``A fine-grained gpu sharing and job
  scheduling for deep learning jobs on the cloud,'' \emph{The Journal of
  Supercomputing}, vol.~81, no.~2, p. 361, 2025.

\bibitem{kim-etal-2021-gpu}
J.~Kim, S.~Ullah, and D.-H. Kim, ``Gpu-based embedded edge server configuration
  and offloading for a neural network service,'' \emph{The Journal of
  Supercomputing}, vol.~77, no.~8, pp. 8593--8621, 2021.

\bibitem{li-etal-2023-easyscale}
M.~Li, W.~Xiao, H.~Yang, B.~Sun, H.~Zhao, S.~Ren, Z.~Luan, X.~Jia, Y.~Liu,
  Y.~Li \emph{et~al.}, ``Easyscale: Elastic training with consistent accuracy
  and improved utilization on gpus,'' in \emph{Proceedings of the International
  Conference for High Performance Computing, Networking, Storage and Analysis},
  2023, pp. 1--14.

\bibitem{arianyan-etal-2025-systematic}
E.~Arianyan, N.~Gholipour, D.~Maleki, N.~Ghorbani, A.~Sepahvand, and
  P.~Goudarzi, ``A systematic review and classification of hpc-related emerging
  computing technologies,'' \emph{Electronics}, vol.~14, no.~12, p. 2476, 2025.

\bibitem{dai-etal-2025-state}
F.~Dai, M.~A. Hossain, and Y.~Wang, ``State of the art in parallel and
  distributed systems: Emerging trends and challenges,'' \emph{Electronics},
  vol.~14, no.~4, p. 677, 2025.

\bibitem{ahmad-etal-2025-containers}
I.~Ahmad, T.~Autto, T.~Das, J.~H{\"a}m{\"a}l{\"a}inen, P.~Jalonen,
  V.~J{\"a}rvinen, H.~Kallio, T.~Kankainen, T.~Kolehmainen, P.~Kontio
  \emph{et~al.}, ``Containers as the quantum leap in software development,''
  \emph{arXiv preprint arXiv:2501.07204}, 2025.

\bibitem{marketgrowthreports2025container}
\BIBentryALTinterwordspacing
{Market Growth Reports}, ``Container orchestration market size, share, growth,
  and industry analysis, by type (platform, services), by application
  (telecommunications and it, bfsi, government and public sector, healthcare,
  retail and consumer goods, manufacturing, others), regional insights and
  forecast to 2035,'' Dec. 2025, last updated 18-Dec-2025. [Online]. Available:
  \url{https://www.marketgrowthreports.com/market-reports/container-orchestration-market-102900}
\BIBentrySTDinterwordspacing

\bibitem{gasser1986integration}
L.~Gasser, ``The integration of computing and routine work,'' \emph{ACM
  Transactions on Information Systems (TOIS)}, vol.~4, no.~3, pp. 205--225,
  1986.

\bibitem{star1994steps}
S.~L. Star and K.~Ruhleder, ``Steps towards an ecology of infrastructure:
  Complex problems in design and access for large-scale collaborative
  systems,'' in \emph{Proceedings of the 1994 ACM Conference on Computer
  Supported Cooperative Work}, ser. CSCW '94.\hskip 1em plus 0.5em minus
  0.4em\relax Chapel Hill, NC, USA: ACM, 1994, pp. 253--264.

\bibitem{orlikowski2007sociomaterial}
W.~J. Orlikowski, ``Sociomaterial practices: Exploring technology at work,''
  \emph{Organization Studies}, vol.~28, no.~9, pp. 1435--1448, 2007.

\bibitem{davis1989perceived}
F.~D. Davis, ``Perceived usefulness, perceived ease of use, and user acceptance
  of information technology,'' \emph{MIS Quarterly}, vol.~13, no.~3, pp.
  319--340, 1989.

\bibitem{davis1989user}
F.~D. Davis, R.~P. Bagozzi, and P.~R. Warshaw, ``User acceptance of computer
  technology: A comparison of two theoretical models,'' \emph{Management
  Science}, vol.~35, no.~8, pp. 982--1003, 1989.

\bibitem{venkatesh2003utaut}
V.~Venkatesh, M.~G. Morris, G.~B. Davis, and F.~D. Davis, ``User acceptance of
  information technology: Toward a unified view,'' \emph{MIS Quarterly},
  vol.~27, no.~3, pp. 425--478, 2003.

\bibitem{venkatesh2012utaut2}
V.~Venkatesh, J.~Y.~L. Thong, and X.~Xu, ``Consumer acceptance and use of
  information technology: Extending the unified theory of acceptance and use of
  technology,'' \emph{MIS Quarterly}, vol.~36, no.~1, pp. 157--178, 2012.

\bibitem{tuunanen2024dealing}
T.~Tuunanen, R.~Winter, and J.~v. Brocke, ``Dealing with complexity in design
  science research: a methodology using design echelons,'' \emph{MIS
  quarterly}, vol.~48, no.~2, pp. 427--458, 2024.

\bibitem{peffers2007design}
K.~Peffers, T.~Tuunanen, M.~A. Rothenberger, and S.~Chatterjee, ``A design
  science research methodology for information systems research,''
  \emph{Journal of management information systems}, vol.~24, no.~3, pp. 45--77,
  2007.

\bibitem{ferreira-campos-2025-exploratory}
F.~Ferreira and J.~Campos, ``An exploratory study on the usage of quantum
  programming languages,'' \emph{Science of Computer Programming}, vol. 240, p.
  103217, 2025.

\bibitem{heikkinen-etal-2025-towards}
R.~Heikkinen, M.~Haghparast, and T.~Mikkonen, ``Towards understanding the
  developer experience in quantum software development,'' in
  \emph{International Conference on Product-Focused Software Process
  Improvement}.\hskip 1em plus 0.5em minus 0.4em\relax Springer, 2025, pp.
  533--542.

\bibitem{matthews-2021-get}
D.~Matthews, ``How to get started in quantum computing,'' \emph{Nature}, vol.
  591, no. 7848, pp. 166--167, 2021.

\bibitem{moller-vuik-2017-impact}
M.~M{\"o}ller and C.~Vuik, ``On the impact of quantum computing technology on
  future developments in high-performance scientific computing,'' \emph{Ethics
  and information technology}, vol.~19, no.~4, pp. 253--269, 2017.

\bibitem{petrosyan-astsatryan-2022-serverless}
D.~Petrosyan and H.~Astsatryan, ``Serverless high-performance computing over
  cloud,'' \emph{Cybernetics and Information Technologies}, vol.~22, no.~3, pp.
  82--92, 2022.

\bibitem{khan-etal-2023-software}
A.~A. Khan, A.~Ahmad, M.~Waseem, P.~Liang, M.~Fahmideh, T.~Mikkonen, and
  P.~Abrahamsson, ``Software architecture for quantum computing systems—a
  systematic review,'' \emph{Journal of Systems and Software}, vol. 201, p.
  111682, 2023.

\bibitem{petrosyan-2025-scheduling}
D.~Petrosyan, ``Scheduling ml and hpc jobs with shoc platform over
  kubernetes,'' \emph{Physics of Particles and Nuclei}, vol.~56, no.~6, pp.
  1581--1585, 2025.

\bibitem{zhou-etal-2023-orchestration}
N.~Zhou, G.~Scorzelli, J.~Luettgau, R.~R. Kancharla, J.~J. Kane, R.~Wheeler,
  B.~P. Croom, P.~Newell, V.~Pascucci, and M.~Taufer, ``Orchestration of
  materials science workflows for heterogeneous resources at large scale,''
  \emph{The International Journal of High Performance Computing Applications},
  vol.~37, no. 3-4, pp. 260--271, 2023.

\bibitem{khan-etal-2025-advancing}
A.~Khan, D.~Taibi, C.~M.~Perrault, and M.~A. Akbar, ``Advancing quantum
  software engineering: A vision of hybrid full-stack iterative model,'' in
  \emph{Proceedings of the 40th ACM/SIGAPP Symposium on Applied Computing},
  2025, pp. 1444--1448.

\bibitem{awan-etal-2022-quantum}
U.~Awan, L.~Hannola, A.~Tandon, R.~K. Goyal, and A.~Dhir, ``Quantum computing
  challenges in the software industry. a fuzzy ahp-based approach,''
  \emph{Information and software technology}, vol. 147, p. 106896, 2022.

\bibitem{juarez-ramirez-etal-2024-skills}
R.~Ju{\'a}rez-Ram{\'\i}rez, S.~Jimenez, C.~X. Navarro, C.~Guerra-Garc{\'\i}a,
  H.~G. Perez-Gonzalez, C.~Fernandez-y Fernandez, J.~Ortiz-Hern{\'a}ndez, and
  K.~Cancino, ``Skills required for quantum computing: A comprehensive review
  of recent studies,'' \emph{Programming and Computer Software}, vol.~50,
  no.~8, pp. 844--874, 2024.

\bibitem{hekkert2003most}
P.~Hekkert, D.~Snelders, and P.~C. Van~Wieringen, ``‘most advanced, yet
  acceptable’: Typicality and novelty as joint predictors of aesthetic
  preference in industrial design,'' \emph{British journal of Psychology},
  vol.~94, no.~1, pp. 111--124, 2003.

\bibitem{purohit2024building}
A.~Purohit, M.~Kaur, Z.~C. Seskir, M.~T. Posner, and A.~Venegas-Gomez,
  ``Building a quantum-ready ecosystem,'' \emph{IET Quantum Communication},
  vol.~5, no.~1, pp. 1--18, 2024.

\bibitem{teitsma2025quantum}
M.~Teitsma, I.~Ahmed, and J.~van Velzen, ``Quantum organisational readiness
  levels,'' 2025.

\bibitem{silberer2023}
J.~Silberer, S.~Astfalk, P.~Planing, and P.~M{\"u}ller, ``User needs over time:
  The market and technology maturity model (mtmm),'' \emph{Journal of
  Innovation and Entrepreneurship}, vol.~12, no.~1, p.~39, 2023.

\bibitem{herrmann2023quantumutility}
N.~Herrmann, D.~Arya, M.~W. Doherty, A.~Mingare, J.~C. Pillay, F.~Preis, and
  S.~Prestel, ``Quantum utility -- definition and assessment of a practical
  quantum advantage,'' in \emph{2023 IEEE International Conference on Quantum
  Software (QSW)}.\hskip 1em plus 0.5em minus 0.4em\relax IEEE, 2023, pp.
  162--174.

\bibitem{bova2023quantumeconomicadvantage}
F.~Bova, A.~Goldfarb, and R.~G. Melko, ``Quantum economic advantage,''
  \emph{Management Science}, vol.~69, no.~2, pp. 1116--1126, 2023.

\bibitem{townes2026gtrl}
M.~S. Townes, ``A generalized technology readiness level scale for measuring
  technology maturity: Development and pilot validation study,'' \emph{Journal
  of Research Management and Administration}, vol.~5, no.~1, p. 2026011801,
  2026.

\bibitem{stirbu-etal-2024-qubernetes}
V.~Stirbu, O.~Kinanen, M.~Haghparast, and T.~Mikkonen, ``Qubernetes: Towards a
  unified cloud-native execution platform for hybrid classic-quantum
  computing,'' \emph{Information and Software Technology}, vol. 175, p. 107529,
  2024.

\bibitem{cipollini2026three}
M.~Cipollini, S.~Rizzo, S.~Iserte, P.~Viviani, G.~Vitali, M.~Barbieri,
  G.~Bettonte, E.~Boella, F.~Ganz, R.~Rocco \emph{et~al.}, ``Three ways to
  share a qpu: Scheduling strategies for hybrid quantum-hpc applications,''
  \emph{arXiv preprint arXiv:2604.14955}, 2026.

\bibitem{pehlivanoglu2026qurator}
S.~Pehlivanoglu, U.~de~Muelenaere, P.~Kogge, and A.~Sabry, ``Qurator:
  Scheduling hybrid quantum-classical workflows across heterogeneous cloud
  providers,'' \emph{arXiv preprint arXiv:2604.05505}, 2026.

\bibitem{cacheiro2025qmiotightlyintegratedhybrid}
\BIBentryALTinterwordspacing
J.~Cacheiro, Álvaro C~Sánchez, R.~Rundle, G.~B. Long, G.~Dold, J.~Friel, and
  A.~Gómez, ``Qmio: A tightly integrated hybrid hpcqc system,'' 2025.
  [Online]. Available: \url{https://arxiv.org/abs/2505.19267}
\BIBentrySTDinterwordspacing

\bibitem{wohlin2012experimentation}
C.~Wohlin, P.~Runeson, M.~H{\"o}st, M.~C. Ohlsson, B.~Regnell, A.~Wessl{\'e}n
  \emph{et~al.}, \emph{Experimentation in software engineering}.\hskip 1em plus
  0.5em minus 0.4em\relax Springer, 2012, vol. 236.

\end{thebibliography}

\end{document}